# Temperature dependence of spin-orbit torque effective fields in the diluted magnetic semiconductor (Ga,Mn)As

B. Howells, K.W. Edmonds, R.P. Campion, and B.L. Gallagher

*School of Physics and Astronomy, University of Nottingham, Nottingham NG7 2RD, United Kingdom*

We report on a study of the temperature-dependence of current-induced effective magnetic fields due to spin-orbit interactions in the diluted ferromagnetic semiconductor (Ga,Mn)As. Contributions from the effective fields as well as from the anomalous Nernst effect are evident in the difference between transverse resistance measurements as a function of an external magnetic field for opposite orientations of the applied current. We separately extract these contributions by fitting to a model of coherently rotating magnetization. The component of the effective field with Dresselhaus symmetry is substantially enhanced with increasing temperature, while no significant temperature-dependence is observed for the component with Rashba symmetry.

Control of ferromagnetism using electrical current has been widely studied for both fundamental interest and commercial exploitation. Spin transfer torque, in which a spin-polarized electrical current induces magnetization switching in a layered magnetic structure, offers a route to high speed, low power memory applications.[1] More recently, it has been shown that a combination of spin-orbit coupling and broken inversion symmetry can result in a magnetic torque even in a uniformly magnetized material. This effect, first reported for the diluted ferromagnetic semiconductor (Ga,Mn)As,[2] has since been demonstrated at room temperature in metallic ferromagnetic heterostructures,[3] enabling electrically driven magnetic switching in non-volatile memory devices. However, there has been considerable debate about the nature of the spin-orbit-induced torque in the latter structures, with spin accumulation due to the spin Hall effect also playing a role.[4,5] Also, in (Ga,Mn)As, certain aspects of the observed effect are not well understood, including its strong temperature-dependence.[6]

The spin-orbit torque induced by an electric current can be described as an effective magnetic field, in a direction which depends on the direction of the applied current. In epitaxial (Ga,Mn)As films, spin-orbit torques result from broken local symmetry within the bulk of the layers, where biaxial and shear strains lead to effective fields with the symmetry of linear Dresselhaus and Rashba spin-orbit interactions respectively (see Fig. 1(a)).[2,7,8] This situation is distinct from the metallic ferromagnetic heterostructures where the role of interfaces is crucial. Two experimental techniques have so far been used to characterize the current-induced fields in (Ga,Mn)As. In the first of these,[2,6,8] a non-saturating magnetic field is rotated in the plane of the sample, and switching of magnetic domains between low-energy orientations is monitored using the planar Hall effect (PHE). The magnitude of the current-induced field can be inferred from the dependence of the switching angle on the size and direction of the current. The second technique[7,9] utilizes resonant excitation of the magnetization by a microwave-frequency current. The resulting magnetization precession is detected as an anisotropic magnetoresistance (AMR), and by determining its dependence on an external magnetic field the magnitude and orientations of the current-induced fields can be obtained. Studies using both techniques have yielded a current-induced field of order 1 Oe per MA per cm$^2$. By directing the current along different crystalline axes, the Dresselhaus and Rashba style effective fields $H_D$ and $H_R$ can be separated, and measurements to date have indicated that the ratio of $H_D$ to $H_R$ in (Ga,Mn)As is around 3:1.[7,8] Intriguingly, a study of the temperature-dependence indicated that the total effective field increased by around a factor of 3 between 30K and 110K.[6] However, since the magnetic anisotropy fields and domain pinning fields vary strongly over the same temperature range, it is not clear to what extent this represents a real temperature-dependence of the current-induced fields.

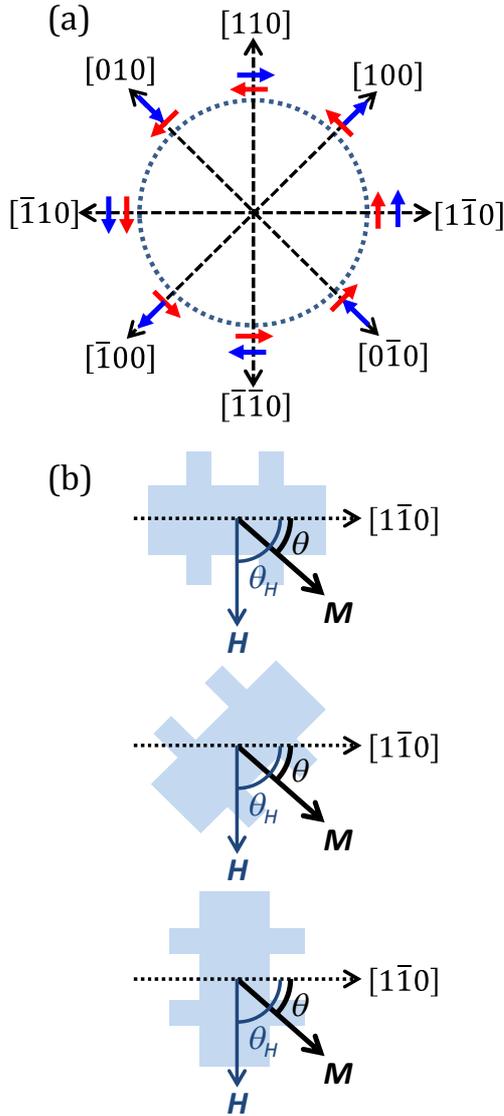

Fig. 1(a). Orientations of the current-induced effective magnetic fields due to the linear Dresselhaus (blue) and Rashba (red) spin-orbit interactions, for current in the (001) plane. (b) Schematic of the Hall bar samples and definition of the angles $\theta$ and $\theta_H$.

Here we report on detailed measurements of the current-induced fields in a (Ga,Mn)As thin film using a different technique. The PHE is recorded under sweeping (rather than rotating) magnetic field for different applied current directions, and the current-induced fields are obtained in a regime of coherent magnetization rotation. A contribution to the transverse voltage due to the anomalous Nernst effect under a vertical temperature gradient is accounted for in the analysis. Our measurements confirm that the effective field increases with increasing temperature, and also show that the part of the effective field with Dresselhaus symmetry dominates its temperature-dependence.

The sample studied is a 25nm (Ga,Mn)As film, grown by molecular beam epitaxy on GaAs(001) and annealed in air at 180°C for 48 hours.[10] The (Ga,Mn)As film is under compressive strain which leads to a strong in-plane magnetic anisotropy.[11] The layer has a nominal Mn concentration of 12% and a Curie temperature of 170K. Low-temperature and high magnetic field Hall measurements of similarly grown and annealed samples indicated a carrier density of around $1.3 \times 10^{21}$ cm$^{-3}$.[12] The sample was patterned into Hall bars of width of 10μm and length between voltage contacts of 10μm, with current channels along the [100], [110] and [1$\bar{1}$0] crystal directions (Fig. 1(b)).

The measurements were performed as follows. At a given temperature, a constant *DC* current density of 1.26MA/cm$^2$ was applied along each Hall bar. The sample temperature was allowed to stabilize under this high current density, for which significant (10-20K) Joule heating occurred. The transverse voltage $V_{xy}$ was then recorded as an external magnetic field was swept from -4kOe to +4kOe. The current direction was reversed, and the magnetic field sweep was repeated. For each Hall bar, the measurement was performed for external magnetic fields applied in the film plane and at 0°, 45°, 90° and 135° to the direction of the Hall bar channel. The temperature of the sample during the measurement was determined from its longitudinal resistance $R_{xx}$, which was calibrated using the $R_{xx}$ versus temperature curve obtained for low current densities.

The transverse resistance due to the PHE in (Ga,Mn)As is given by $R_{xy} = C\sin 2\phi$, where $\phi$ is the angle between the current and magnetization directions, and $C$ is a temperature-dependent amplitude that depends on the orientation of the current with respect to the principal crystal axes.[13,14] In this paper, we utilize the PHE for different orientations of the current and external magnetic field, in order to determine the magnitude and direction of the current-induced effective magnetic fields. For example, for current $I$ and magnetization **M** both aligned along the $[1\bar{1}0]$ easy axis, a current-induced effective field along the $[110]$ direction will cause a tilt in **M**, resulting in the appearance of a planar Hall resistance that depends on the relative sizes of the effective field and the magnetic anisotropy fields. Reversing the current direction will reverse the direction of the effective field and the sign of the planar Hall resistance. Also, applying a magnetic field along the current direction will reduce the tilt of the magnetization, decreasing the magnitude of the planar Hall resistance.

There is another contribution to the transverse voltage $V_{xy}$ that depends, indirectly, on the magnitude of the current, which must also be considered. Due to Joule heating, carrier holes within the film are out of thermal equilibrium with the lattice. The top interface of the (Ga,Mn)As film is with vacuum, while the bottom interface is with the GaAs substrate. There is therefore a net heat flow perpendicular to the film and a corresponding vertical temperature gradient $\nabla T$ within the film. As a result, an electric field will be induced in the direction of **M** × $\nabla T$, due to the anomalous Nernst effect (ANE). Therefore, when **M** has a component directed along the current channel of a Hall bar, a thermoelectric voltage will appear between the transverse voltage contacts. The thermoelectric voltage due to the ANE, $V_{ANE}$, will vary with $\cos\theta$, and its contribution to the measured $R_{xy}=V_{xy}/I$ will change sign on reversing the direction of the current.

Figure 2 shows $R_{xy}$ versus external magnetic field for the three studied Hall bars, for magnetic fields applied in the plane of the Hall bar and parallel, perpendicular and at 45° to the current direction. In Fig. 2(a-c) the average $R_{xy}$ for opposite orientations of the current is shown, while Fig. 2(d-f) shows the difference $\Delta R_{xy}$ between measurements for positive and negative currents. The current-averaged $R_{xy}$ shows the expected behavior for a system with a dominant uniaxial magnetic anisotropy along the $[1\bar{1}0]$ crystal axis: the $R_{xy}$ shows little variation when the magnetic field is applied along this axis, as the magnetization remains fixed along an easy orientation. Small asymmetries in the current-averaged $R_{xy}$ versus magnetic field plots are due to a small unintentional misalignment of the magnetic field with the film plane. A sizeable $\Delta R_{xy}$ is observed for certain orientations of the current and external magnetic field. At low fields, behavior associated with a tilting of the magnetization can be observed. At high fields the $\cos\theta$ dependence due to the anomalous Nernst voltage is clearly evident: the measured voltage is close to zero for M perpendicular to $I$, a maximum / minimum for M parallel/antiparallel to $I$, and ~$1/\sqrt{2}$ of the maximum / minimum value for $\theta = 45^0 / 135^0$.

The $R_{xy}$ and $\Delta R_{xy}$ versus field plots can be described by minimizing the magnetic free energy $U$ of the (Ga,Mn)As film, modelled using the equation:

$$U = H_U\sin^2\theta - (H_C/4)\sin^2 2\theta - H_{ext}\cos(\theta - \theta_H) - H_{eff1}\cos(\theta - \theta_I - \pi/2) - H_{eff2}\cos(\theta - \theta_I) \quad (1)$$

where $H_U$ and $H_C$ are the in-plane uniaxial and cubic magnetic anisotropy fields, $H_{ext}$ is the external magnetic field, $H_{eff1}$ and $H_{eff2}$ are the current-induced effective fields perpendicular and parallel to the current directions, and $\theta$, $\theta_H$ and $\theta_I$ are the angles of the magnetization, magnetic field and current with respect to the $[1\bar{1}0]$ uniaxial easy axis. The transverse resistance is then given by

$$R_{xy} = C\sin(2\theta - 2\theta_I) + (V_{ANE}/I)\cos(\theta - \theta_I) + R_{mis} \quad (2)$$

where $C$ is the amplitude of the planar Hall resistance, $V_{ANE}$ is the thermoelectric voltage due to the anomalous Nernst effect, $I$ is the current and $R_{mis}$ is an offset due to a small misalignment of voltage probes. We emphasize that the effective fields $H_{eff1}$ and $H_{eff2}$ both arise within the bulk of the (Ga,Mn)As layers; due to the insulating substrate, there is no parallel conduction, and torques due to, for example, spin Hall effect in neighboring layers are not present.

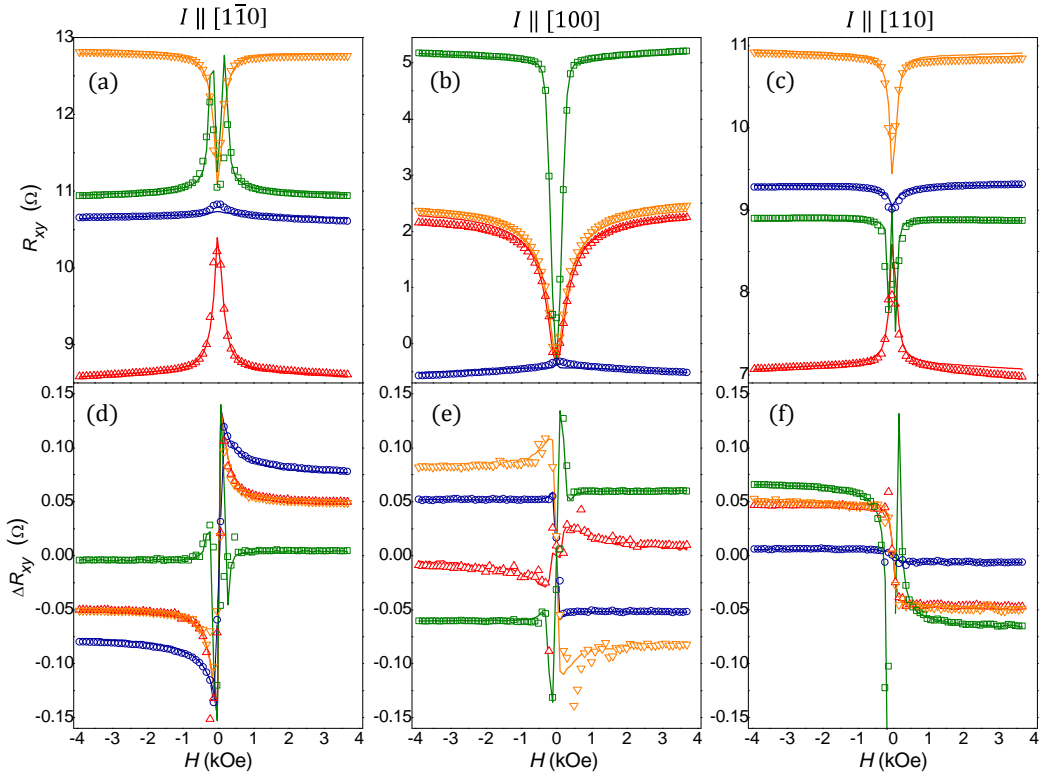

*Fig. 2. Average (a-c) and difference (d-f) of the transverse resistance $R_{xy}$ versus external magnetic field measured for opposite directions of the current density J=1.26MA/cm$^2$, at a sample temperature of 101K. The current is oriented along the $[1\bar{1}0]$ in (a, d), [100] in (b, e), and [110] in (c, f). Blue circles are for magnetic field along $[1\bar{1}0]$, green squares [110], red up triangles [010], and orange down triangles [100].*

Firstly, the current-averaged $R_{xy}$ was modelled using equations (1) and (2), to obtain the anisotropy fields $H_U$ and $H_C$. The $\Delta R_{xy}$ plots were then fitted to obtain the current-induced effective fields and the anomalous Nernst voltage. For current along the [110] and $[1\bar{1}0]$ directions, $H_{eff2}$ was set to zero due to the expected symmetry of $H_D$ and $H_R$, while for current along [100], both $H_{eff1}$ and $H_{eff2}$ were used as free fitting parameters. The fitting procedure assumes that the magnetization remains single-domain-like throughout the field sweep, which is valid away from $H_{ext} \approx 0$, where domain switching events occur.[13] It also neglects possible variations in the size of the current-induced effective fields on the angle between magnetization and current, or out-of-plane components of the current-induced fields.[9,15,16]

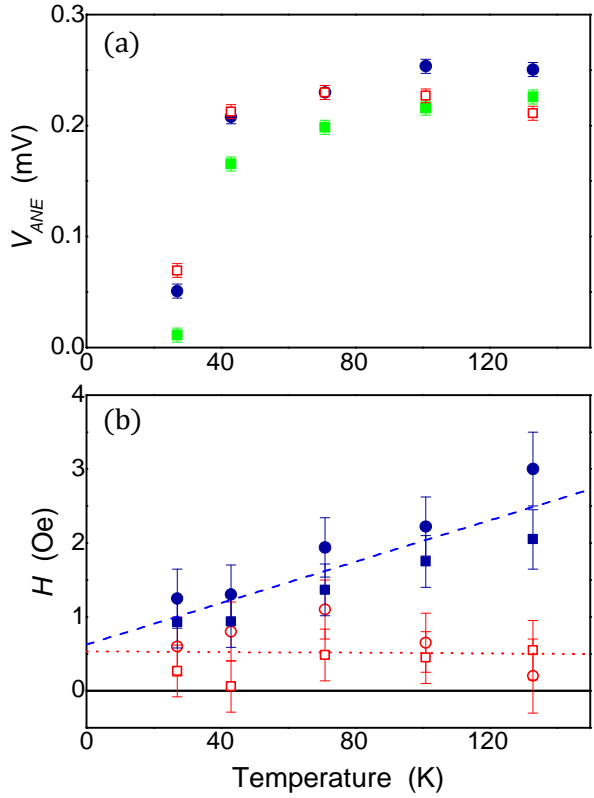

*Fig. 3(a). Anomalous Nernst voltage versus temperature for the Hall bars oriented along the [100] (blue circles), [1$\bar{1}$0] (red open squares) and [110] (green filled squares) crystal directions. (b) Current-induced effective magnetic fields versus temperature. Open and filled symbols represent the Rashba and Dresselhaus fields, $H_R$ and $H_D$. Circles were obtained for the [100] Hall bar, and squares were obtained from the [110]/[1$\bar{1}$0] Hall bars. Dotted and dashed lines are linear fits to the $H_R$ and $H_D$ data, to indicate the trends.*

The in-plane magnetic anisotropy fields given by the fitting procedure are consistent with the values obtained from pump-probe magneto-optical measurements of similar samples,[11] with $H_U$ around 200Oe at the lowest temperature studied, and $H_C$ around an order of magnitude smaller. Figure 3 plots the obtained values of the anomalous Nernst voltage and the current-induced effective fields as a function of temperature. The fields with Rashba and Dresselhaus symmetry were obtained by linear combinations of $H_{eff1}$ for the [110] and [1$\bar{1}$0] oriented Hall bars, and also directly from $H_{eff1}$ and $H_{eff2}$ for the [100] oriented Hall bar. The error bars include the effect of small uncertainties in the alignment of $\theta_H$ and $\theta_I$ of $\pm 5°$ and $\pm 2°$ respectively.

The anomalous Nernst voltages, shown in Fig. 3(a), display a comparable temperature-dependence to previous studies of annealed (Ga,Mn)As films,[17] and are strongly suppressed at the lowest temperature studied. Averaged over all temperatures measured, $H_D$ and $H_R$ are 1.7Oe and 0.5Oe at a current density $J$ of 1.26MA/cm$^2$, giving a comparable magnitude of $H_{eff}/J$ as well as $H_D/H_R$ to previous studies.[2,6,7,8] Also, although there is significant uncertainty in the values of the current-induced effective fields shown in Fig. 3(b), there is a clear systematic increase with increasing temperature in the part of the field with Dresselhaus symmetry, for both the [100] and [110]/[1$\bar{1}$0] Hall bars.

The observed increase with temperature of the current-induced effective magnetic field confirms previous observations, which were based on measurements of shifts in the magnetization orientation inferred from planar Hall measurements in a rotating magnetic field.[6] Our measurements, although subject to greater uncertainty, are less sensitive to variations in the temperature-dependence of the anisotropy fields, which are accounted for in the fitting procedure. They also suggest a different temperature dependence for $H_D$ and $H_R$, with the latter showing no significant variation over the temperature range studied within the experimental uncertainty.

Calculations of the linear Dresselhaus spin-orbit torque in III-V magnetic semiconductors described a non-linear dependence on the *p-d* exchange splitting, due to the interplay of magnetic and non-magnetic scattering.[16] Since the exchange splitting is strongest at low temperatures, this may be related to our observed temperature-dependence. Ref. [16] also demonstrated that the effective fields may show a significant dependence on the angle of the magnetization, due to the warped Fermi surface. We obtain good fits to our measured data shown in Fig. 2 without including such anisotropy effects, although we cannot rule them out. Indeed, the values of $H_D$ obtained from the [110]/[1$\bar{1}$0] Hall bars are systematically lower than from the [100] Hall bar, although the differences are comparable to the experimental uncertainty.

In summary, we have extracted current-induced magnetic fields in (Ga,Mn)As, from planar Hall effect measurements and a minimized free energy fitting procedure. The obtained magnitudes of the effective fields with linear Dresselhaus and Rashba symmetry are consistent with previous experiments. The Dresselhaus field shows a significant temperature-dependence while the Rashba field is (within error) temperature-independent. These results may offer further insight into the factors controlling the magnitudes and symmetries of current-induced magnetic fields in magnetic materials, which are of increasing interest for magnetic memory applications.

This work was supported by the UK EPSRC through grant EP/H002294/1. We thank Tomas Jungwirth and Joerg Wunderlich for useful discussions.